# Self-heating effects of the surface oxidized FeCo nanoparticles colloid under alternating magnetic field

Jinu Kim[1], Baekil Nam[2] and Ki Hyeon Kim[1]

[1]Department of Physics, Yeungnam University, Gyeongsan, 38541 Korea

[2]School of General Education, Yeungnam University, Gyeongsan, 38541 Korea

**Abstract**

To evaluate the self-heating effects of FeCo magnetic nanoparticles, the surface oxidized FeCo nanoparticles were synthesized by co-precipitation method with the reduction reaction without any post treatments. As-synthesized FeCo nanoparticles exhibited the mean diameter of about 39 nm with the oxidized shell thickness of about 4-5 nm. The saturation magnetization and coercivity were obtained 172 emu/g and 268 Oe at 300 K, respectively. The heat elevation of the FeCo magnetic colloid was measured under alternating magnetic fields of 76, 102, and 127 Oe with selectable frequencies of 190, 250 and 355 kHz. The heat temperature increased up to about 45 $^{o}$C from initial temperature of 24 $^{o}$C under 127 Oe and 355 kHz, which the specific absorption exhibited about 35.7 W/g.

Corresponding author: Ki Hyeon Kim

Email: kee1@ynu.ac.kr

## I. Introduction

Magnetic nanoparticles (MNPs) with high magnetic moment are very important in electronic devices and biosensing and biomedical applications such as high density recording, induction core, electromagnetic interference (EMI) materials, magnetic resonance imaging (MRI) contrast agent and drug delivery etc. Among these magnetic materials, FeCo MNPs are one of the most important magnetic materials in their applications [1-6]. FeCo MNPs can be widely synthesized by a range of methods which are sol-gel, thermal decomposition, reduction of metal salts, ultrasonic-assisted wet chemical route and co-precipitation [6-17]. Among these chemical methods, a chemical co-precipitation method has many advantages, such as a simple process, a high material quality with a control of the fine particle size and relatively a rapid reaction time at room temperature [8]. During the co-precipitation process, metal salts can be reduced to form nanoparticles using a wide range of common reducing agents, including hydrazine, sodium borohydride and lithium borohydride. On the other hand, this method requires surfactants for shape control and mono-dispersion. Moreover, boron was also reported to be incorporated into the reduced metals using borohydride-based reducing agents [17-19]. B. E. Yong et. al. [7] reported the co-precipitation synthesis of 35 nm and 12 nm MNPs possessing a saturation magnetization of 148 emu/g and 23 emu/g, respectively. Kang et. al. [18] reported the fabrication of 8 nm - 22 nm FeCo MNPs with a maximum saturation magnetization of 147.8 emu/g (as-synthesized) and 225.4 emu/g (annealed in the range of 400-800 °C). When the FeCo MNPs were generated by an alternating magnetic field (AMF), the heat from magnetic loss occurs through a combination of eddy currents, hysteresis losses, and relaxation losses. The eddy current loss is governed the skin depth of materials which is defined by the applied frequency, conductivity and permeability of materials. However, the eddy current loss can be negligible in a nano-sized particles within a few hundred kHz region. The magnetic loss in the agglomerated MNPs is mainly caused by the hysteresis loss due to the inter-particle interaction [20] although the dispersed MNPs is mainly generated by Neel relaxation due to the magnetic spins in single-domain MNPs. [21-23] The MNPs movement in the suspending medium contributes to the heat loss due to the interaction between a thermal force and viscous drag, which can be estimated by the Brownian relaxation. These total losses of the MNPs are governed by the strength of the AMF and their operating frequencies as well as the intrinsic magnetic properties. [24] Thus, we synthesized FeCo MNPs and evaluated the self-heating effects of the surface oxidized FeCo MNPs colloid under alternating magnetic field.

## II. Experimental

The FeCo MNPs were synthesized using a co-precipitation method without any post treatments. Cobalt chloride hexahydrate (CoCl$_2$·H$_2$O) and iron chloride hexahydrate (FeCl$_3$·6H$_2$O) were dissolved into 200 mL of deionized water for 10 min. under a N$_2$ environment at 70 °C to avoid the unexpected oxidation. Sodium borohydride (NaBH$_4$, 1.5 mol) as a reducing agent was dropped into the metal solution using a dropping funnel and then reacted with metallic salts during 60 min at 70 °C with pH values of approximately 8.9. The reaction of the borohydride reduction is generally written as follows. [19].

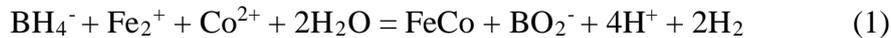
$$BH_4^- + Fe^{2+} + Co^{2+} + 2H_2O = FeCo + BO_2^- + 4H^+ + 2H_2 \quad (1)$$

To remove the residual reducing agent after the reactions, the precipitated FeCo MNPs were washed by deionized water and ethyl alcohol. And then the particles were dried in a vacuum oven at 50 °C for 12 hrs. The structural phase and crystallite size were examined by X-ray diffraction (XRD; X'Pert PRO, PANalytical) using Cu Kα radiation (1.54 Å). Transmission electron microscopy (TEM, JEOL JEM-2200FS) with electron energy dispersive X-ray spectroscopy (EDS) and Energy-filtered transmission electron microscopy (EFTEM) was used to confirm the structure, and morphology of the. The magnetic properties were measured by using Vibrating Sample Magnetometer (VSM, Lakeshore 7410). To evaluate the heat behaviors, the magnetic colloid was prepared with FeCo NPs in water of 4.5 mg/ml. The induction heat temperature were measured by AC heating system which is composed of RF power supply (AMERITHERM INC., HOTSHOT 2.4 kW) and 5.5 turns-helical shaped Cu coil with the 80 mm-height and 70 mm-inner diameter as an AC magnetic field applicator as shown in Fig. 1. The strength of AMF was measured by a magnetic field transducer (SENIS AG.). The temperature was measured by using 4 channel thermometer (FISO, TMI4) and fiber optic probes (FISO, FOT-L-BA) with the resolution and accuracy of 0.1°C and ± 1°C, respectively

## III. Results and Discussion

The representative α-FeCo (110), (200), and (211) peaks were observed by XRD as shown in Fig. 2. The calculated d-spacing of the α-FeCo (110) peaks was about 0.202 nm, indicating a lattice parameter of 2.86 Å, which is similar with that of $Fe_{70}Co_{30}$ bulk material (0.2023 nm and 0.2024 nm; JCPDS 48-1816 and JCPDS 48-1817). The distribution of the synthesized FeCo MNPs mostly exhibited the aggregated chain-like shapes as shown in Fig. 3(a). The α-FeCo with body centered cubic (bcc) phase was confirmed by HRTEM results in Fig. 3(b), which is correspond with that of XRD results. The mean crystalline size exhibited about 40 nm by Scherrer's relation, which is comparable to that of the TEM observation of about 39 nm. And the thickness of surface oxidized layer of FeCo MNPs exhibited around 5nm in Fig. 3(c). The Fe and Co atomic ratio was about 72.5:27.5 as shown in Fig. 3 (d), which compositions followed the expected value of the molar ratio of Fe and Co metal salts. To verify the element distribution in a FeCo MNPs, the EFTEM element mapping images of Fe, Co, and O were shown in Fig. 4, respectively. The surface oxidations and the partially existences of the unexpected FeCo composition are deeply related to the magnetic behavior. The saturation magnetization and coercivity were obtained 172 emu/g and 268 Oe at 300 K, as shown in Fig. 5. The saturation magnetization is about 30 % lower than that of the bulk value ($Fe_{65}Co_{35}$, 245 emu/g). [25] It indicates that the surface oxidation around the FeCo nanoparticles and their canting of the surface moment were attributed to the total magnetic moment. Once the particles have agglomerated, they had the strong magnetic interaction and contributed to the large coercivity with the existence of complex magnetic domains due to the various distribution with different size and shapes of MNPs. To verify the heat elevation of the FeCo MNPs colloid, the specimens were placed in adiabatic jig with the initial temperature of 24 °C. The intensity of AMF and the frequency were generated with 76 Oe, 102 Oe, 127 Oe and 190 kHz, 250 kHz, 355 kHz, respectively. To confirm the reliability of the measuring temperature, the temperature change of reference specimen was simultaneously detected with that of FeCo colloid under same experimental conditions, which reference was prepared the deionized water without FeCo MNPs. Figure 6 (a) shows the heat behaviors of FeCo colloid for the increment of frequency at AMF of 127 Oe. The temperatures were linearly increased with the increment of frequency. The saturated temperatures were exhibited up to about 29, 37.5 and 45 °C with the increment of frequency from 190 kHz to 355 kHz at AMF of 127 Oe, respectively. For the AMF dependence of heat, the elevated temperatures of FeCo colloid were measured under AMF of 76 Oe, 102 Oe and 127 Oe at 190 kHz, respectively, as shown in Fig. 6 (b). The temperatures were slightly increased with the increment of AMF. The saturated temperatures were elevated up to about 24.7, 26.4 and 29.5 °C from the initial temperature of 24°C.

The heat generation by magnetic losses of the agglomerated MNPs is governed by hysteresis loss and interparticle interactions [20] although the dispersed MNPs is almost given by the Brown and Neel relaxations. [23] Especially, the effective magnetic loss of the FeCo MNPs colloid is not easy to predict due to the complicated magnetic interactions with the mixed shapes and sizes by the locally aggregated and partially dispersed MNPs in solvent. Thus, the practical magnetic losses of these FeCo MNPs colloid could be better to evaluate by the experimental results of heat behaviors. In general, the magnetic losses for heat generation can be expressed by the specific loss power (SLP), also called the specific absorption rate (SAR), which is defined as the thermal dissipation per unit of mass of the magnetic material in the presence of an AMF. SLP is equal to $c \times \Delta T/\Delta t \times m_{colloid}/m_{particle}$, where c is the specific heat capacity of sample, $m_{colloid}$ is the mass of colloid, $m_{particle}$ is the mass of the magnetic material in specimen, and $\Delta T/\Delta t$ is the initial slope of the time-temperature heating curve.[26] The initial slope of the time-temperature heating curves were fitted by use of the phenomenological Box-Lucas equation given by $T(t) = A(1-e^{-Bt})$. [27] The fitting parameters *A* and *B* are related to the final temperature and the initial slope of the time-temperature heating ratio. The heat capacity of the system corresponds to $(m_{particle} \cdot c_{particle} + m_{solvent} \cdot c_{solvent}) / (m_{particle} + m_{solvent})$, where $m_{solvent}$ and $c_{solvent}$ are the mass and specific heat capacity of solvent (water, $c_{solvent} = 4.1$ J/Kcm$^3$), respectively. [24], [26] The SAR values were calculated on the measured time-temperature increasing ratio ($\Delta T/\Delta t$) with the change of AMF, frequency, as shown in Fig. 6(c) and (d), respectively. Although the SAR values were not dominant with the change of AMF, these values increased from about 1.24 W/g to 6.7 W/g with the increment of AMF at the fixed frequency of 190 kHz. As increasing the frequency at the fixed AMF of 128 Oe, the SAR increased from about 6.7 W/g to 35.7 W/g, respectively.

## IV. Conclusion

The surface oxidized FeCo MNPs were synthesized and evaluated the self-heating effects of their colloid under AMF. FeCo MNPs colloid showed that the temperatures increased with the increment of AMF, frequency, respectively. The maximum saturated temperature and ΔT showed about 45.3 °C and 21.3 °C and their maximum SAR exhibited about 35.7 W/g at 127 Oe, 355 kHz, respectively. The core-shell structured FeCo MNPs without any post treatment could be employed as one of good candidate materials for bio applications.

**Acknowledgment**

This study was supported by Basic Science Research Program through the National Research Foundation of Korea (NRF) funded by the Ministry of Education, Science and Technology (NRF-2013R1A1A2A 10058888).

**Figure Captions**

Fig. 1. The schematic of alternative magnetic field applicator for heating of MNPs.

Fig. 2. X-ray diffraction patterns of as-synthesized FeCo MNPs.

Fig. 3. TEM images of (a) the aggregated chain-like shaped FeCo MNPs, (b) SAED diffraction pattern, and (c) HRTEM images of core-shell structured FeCo MNPs, (d) the EDS results with compositions of FeCo MNPs, respectively.

Fig. 4. The EFTEM element mapping images of Fe, Co, and O, respectively.

Fig. 5. The magnetization curves of FeCo MNPs.

Fig. 6. The measured temperatures (a)-(b) and SAR (c)-(d) of the FeCo MNPs colloid with increment of frequency (190 to 355 kHz @ 127 Oe) and AMF (76 to 127 Oe @ 190kHz), respectively.

Fig. 1

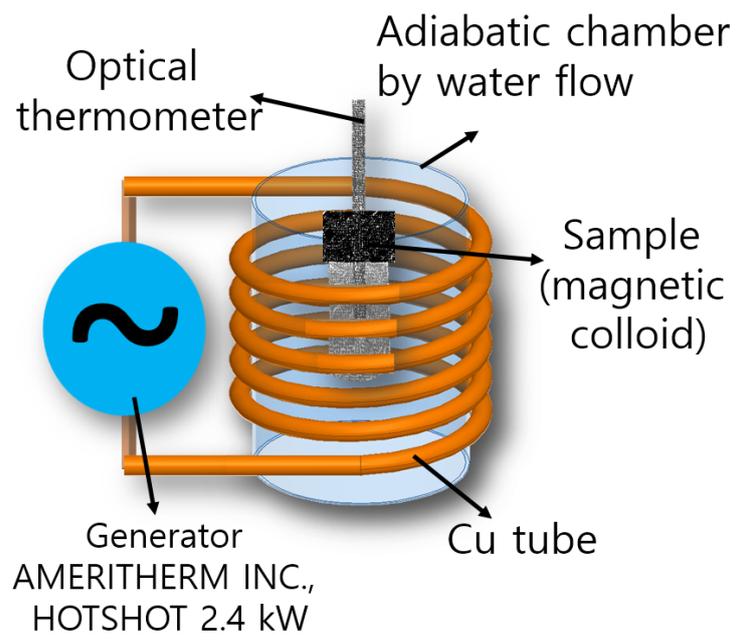

Fig. 2

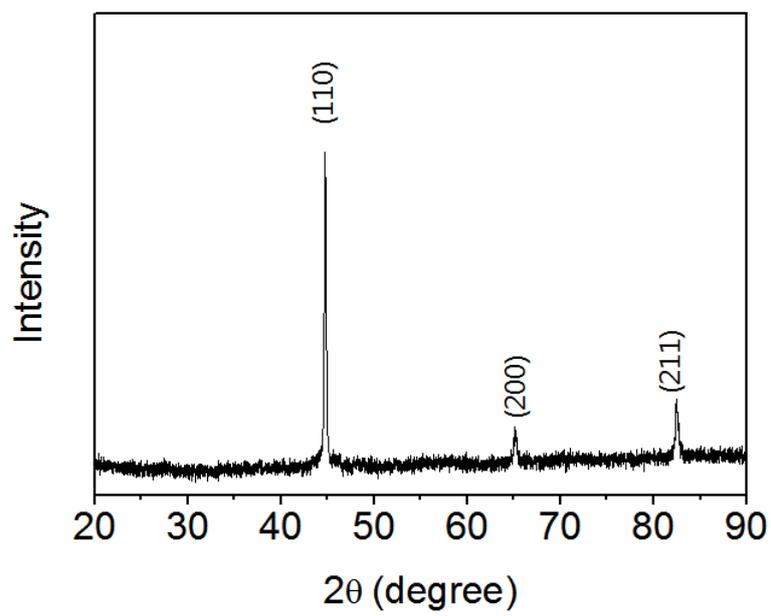

Fig. 3

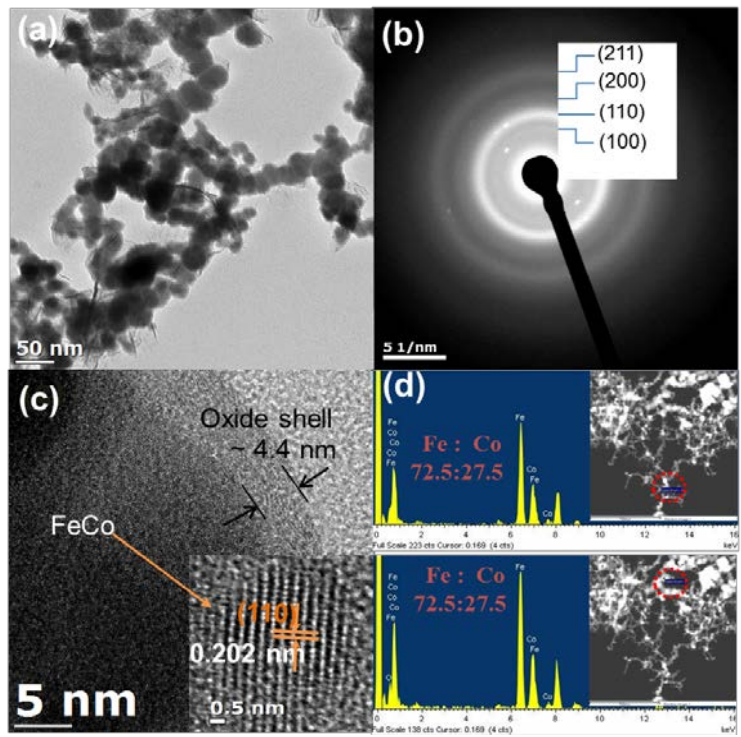

Fig. 4

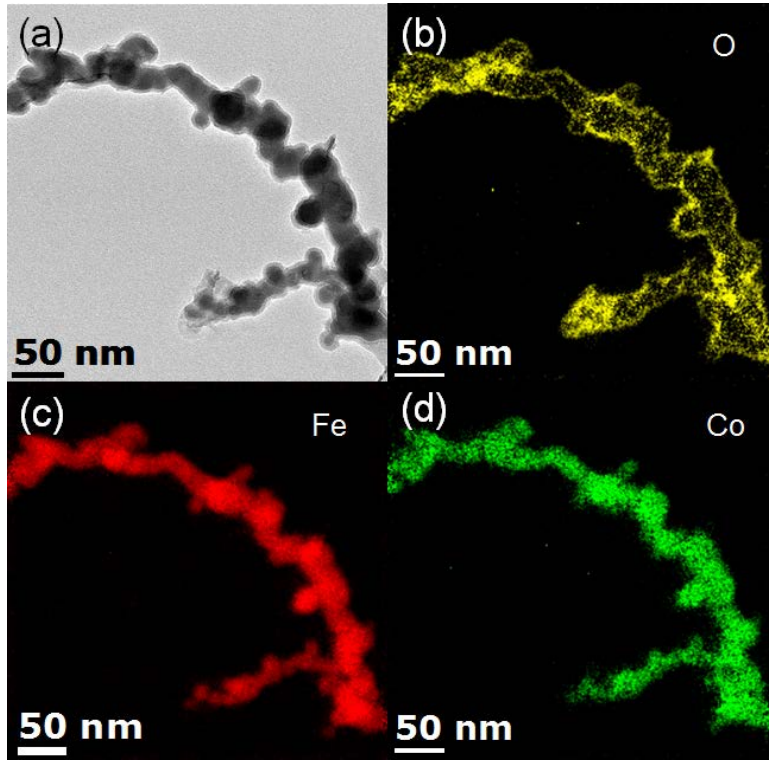

Fig. 5

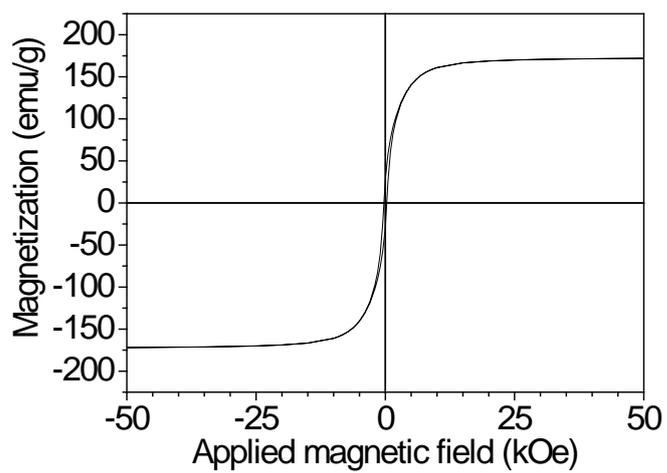

Fig. 6

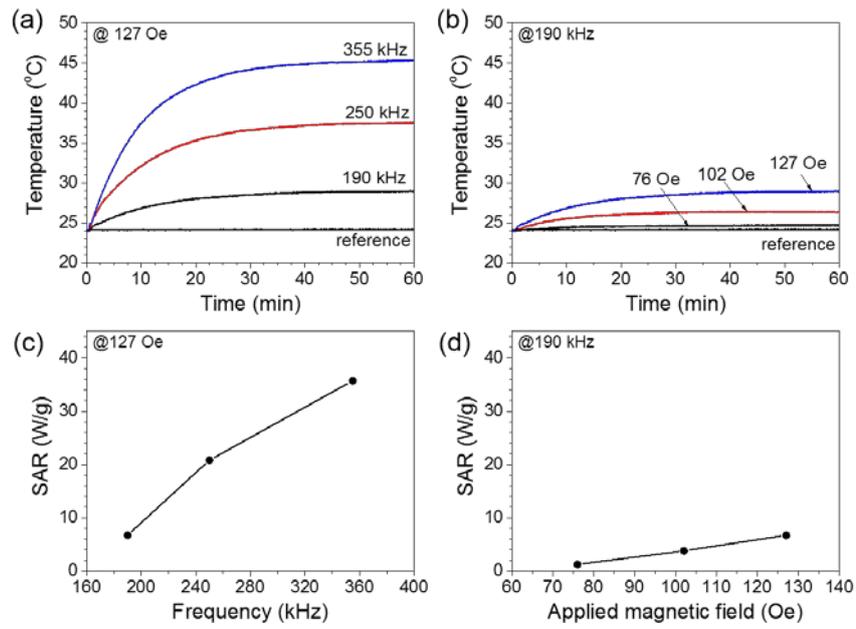